%% file: DiphotonPublication.tex
\pdfoutput=1
\newcommand*{\ATLASLATEXPATH}{latex/}
\documentclass[cernpreprint,texlive=2016,txfonts,UKenglish]{latex/atlasdoc}

\usepackage{\ATLASLATEXPATH atlaspackage}
\usepackage{\ATLASLATEXPATH atlasbiblatex}

\usepackage{\ATLASLATEXPATH atlascontribute}

\usepackage{\ATLASLATEXPATH atlasphysics}

\usepackage{multirow}
\usepackage{enumitem}

\graphicspath{{logos/}{figures/}}

\usepackage{DiphotonPublication-defs}

\input{DiphotonPublication-metadata}
\hypersetup{pdftitle={ATLAS document},pdfauthor={The ATLAS Collaboration}}

\begin{document}

\maketitle

\input{sections/introduction}
\newpage
\input{sections/atlas_detector}
\input{sections/samples}

\input{sections/selection}

\input{sections/signal_modeling}
\input{sections/background_estimates}
\input{sections/statistical_procedure}
\input{sections/results}

\FloatBarrier
\input{sections/conclusion}

\section*{Acknowledgements}

\input{acknowledgements/Acknowledgements}

\clearpage

\printbibliography

\newpage \input{atlas_authlist}

\end{document}

%% file: DiphotonPublication-metadata.tex
\AtlasTitle{Search for new phenomena in high-mass diphoton final states using \intLumiRounded\ of proton--proton collisions collected at $\sqrt{s}=13~\TeV$ with the ATLAS detector}

\author{The ATLAS Collaboration}

\AtlasRefCode{HIGG-2016-17}

\PreprintIdNumber{CERN-EP-2017-132}

\arXivId{1707.04147}

\HepDataRecord{https://www.hepdata.net/record/79924}

\AtlasJournal{Phys.\ Lett.\ B.}
\AtlasJournalRef{\PLB 775 (2017) 105}
\AtlasDOI{10.1016/j.physletb.2017.10.039}

\AtlasAbstract{%
Searches for new phenomena in high-mass diphoton final states with the ATLAS experiment at the
LHC are presented. The analysis is based on $pp$ collision data corresponding to an integrated
luminosity of \intLumi\ at a centre-of-mass energy $\sqrt{s}=13~\TeV$ recorded in 2015 and 2016. Searches are performed for resonances with spin 0, as predicted by theories with an extended Higgs sector, and for resonances with spin 2, using a warped  
extra-dimension model as a benchmark model, as well as for non-resonant signals, assuming a large extra-dimension scenario.
No significant deviation from the Standard Model is observed. 
Upper limits are placed on the production cross section times branching ratio to two photons as a function of the resonance mass.
In addition, lower limits are set on the ultraviolet cutoff scale in the large extra-dimensions model.
}

%% file: sections/introduction.tex
\section{Introduction}
\label{sec:intro}

New high-mass states decaying into two photons are predicted in extensions of the Standard
Model (SM). The diphoton final state provides a clean experimental signature with excellent invariant-mass resolution and moderate backgrounds. 
This Letter presents an update of the searches for new high-mass states decaying into two photons,
using both 2015 and 2016 proton--proton ($pp$) collision datasets recorded at a centre-of-mass energy $\sqrt{s}=13$~\TeV\ by the ATLAS detector at the CERN Large Hadron Collider (LHC), 
corresponding to a total integrated luminosity of \intLumi. The analysis closely follows
that described in Ref.~\cite{Aaboud:2016tru}, and includes small improvements in the photon reconstruction, selection and energy calibration. From the many extensions of the SM that predict new high-mass resonances decaying into two photons,
the two benchmark signal models studied in Ref.~\cite{Aaboud:2016tru} are considered: 
a spin-0 resonance ($X$) as predicted in theories with an extended Higgs sector~\cite{Hill:1987ea,Veltman:1989vw,Binoth:1996au,Schabinger:2005ei,Patt:2006fw,Pruna:2013bma,Lee:1973iz}, and the lightest Kaluza--Klein (KK)~\cite{ModernKK} spin-2
graviton excitation ($G^*$) of a Randall--Sundrum~~\cite{Randall:1999ee} model with one warped extra dimension, later referred to as RS1.
The ATLAS and CMS collaborations reported a modest excess in the diphoton invariant-mass spectra with respect to the SM continuum background near a mass value of 750~\GeV~\cite{Aaboud:2016tru,Khachatryan:2016hje}, using 3.2--3.3~\ifb\ of $pp$ collision data recorded in 2015 at the LHC.
The ATLAS result corresponds to a global significance of 2.1~standard deviations ($\sigma$).
The CMS result corresponds to a global significance of 1.6\,$\sigma$.
No significant excess is observed by CMS in 12.9~\ifb\ of data collected in 2016~\cite{Khachatryan:2016yec}.

In addition to searching for a resonant signal, data are interpreted using the model proposed by Arkani-Hamed, Dimopoulos and Dvali (ADD)~\cite{ArkaniHamed:1998rs}.
Motivated by the weakness of gravity, the ADD model predicts the existence of $n$ extra dimensions of space-time where only gravity can propagate.
If the ultraviolet cutoff scale (\ms) of the Kaluza--Klein spectrum is lower than the Planck scale in the $(4+n)$-dimensional space-time,
the extra dimensions may be detected via virtual KK graviton exchange before being observed via direct KK graviton emission.
The strength of gravity in the presence of extra dimensions is typically parameterized by $ \eta_G = F/\ms^4$,
where $F$ is a dimensionless parameter of order unity reflecting the dependence of the virtual KK graviton exchange on the number of extra dimensions.
Several theoretical formalisms exist in the literature~\cite{Giudice:1998ck,Han:1998sg,Hewett:1998sn}.
While the definition of $\eta_G$ is consistent, each formalism uses a different convention for $F$,
which consequently leads to a different definition of \ms.
The KK graviton exchange creates a set of finely spaced resonances,
which manifests itself as a non-resonant deviation from the expected SM background in the 
diphoton mass distribution due to limited experimental resolution.
The effective diphoton cross section is the result of the SM and ADD amplitudes, as well as their interference. The interference term in the effective cross section is linear in $\eta_G$ and the pure KK graviton exchange term is quadratic in $\eta_G$.
The interference effect is assumed to be constructive in formalisms considered in this Letter.
Previous searches for an ADD graviton signal
in the diphoton decay channel were carried out by the ATLAS~\cite{Aad:2012cy} and CMS~\cite{Chatrchyan:2011fq} experiments based on LHC Run 1 data.
The substantial increase in the centre-of-mass energy in LHC Run 2 greatly enhances the sensitivity to the ADD scenario at higher mass scales~\cite{CMS-EXO-15-009}.

%% file: sections/atlas_detector.tex
\section{ATLAS detector}
\label{sec:detector}

The ATLAS detector~\cite{PERF-2007-01, ATLAS-TDR-19} is a multi-purpose detector with a forward--backward symmetric
cylindrical geometry\footnote{The ATLAS experiment uses a right-handed coordinate system with its origin at the nominal interaction point (IP) in the centre of the detector and the $z$-axis along the beam pipe. The $x$-axis points from the IP to the centre of the LHC ring, and the $y$-axis points upward. Cylindrical coordinates $(r,\phi)$ are used in the transverse plane, $\phi$ being the azimuthal angle around the $z$-axis. The pseudorapidity is defined in terms of the polar angle $\theta$ as $\eta=-\ln\tan(\theta/2)$. The transverse energy
is defined as $\et = E \sin(\theta)$.}. The systems most relevant for the presented searches are the inner detector (ID), immersed in a 2~T
magnetic field produced by a thin superconducting solenoid, and the calorimeters.
The ID consists of fine-granularity silicon pixel and microstrip detectors covering
the pseudorapidity range $|\eta|<2.5$ complemented by a gas-filled straw-tube transition radiation tracker (TRT) at larger radii which covers the region $|\eta| < 2.0$ and provides electron identification capabilities. The electromagnetic (EM) calorimeter is a lead/liquid-argon sampling
calorimeter with accordion geometry. It is divided into a barrel section covering
$|\eta|<1.475$ and two end-cap sections covering $1.375<|\eta|<3.2$. For $|\eta|<2.5$, it is divided
into three layers in depth, which are finely segmented in $\eta$ and $\phi$. A thin presampler
layer, covering $|\eta|<1.8$, is used to correct for fluctuations in upstream energy losses.
Hadronic calorimetry in the region $|\eta|<1.7$ uses steel absorbers and scintillator tiles as the
active medium.  Liquid-argon calorimetry with copper absorbers is used in the hadronic end-cap
calorimeters, which cover the region $1.5<|\eta|<3.2$.  A forward calorimeter using copper and
tungsten absorbers with liquid argon completes the calorimeter coverage up to $|\eta|=4.9$.
The muon spectrometer, located beyond the calorimeters,
consists of three large air-core superconducting
toroid systems with precision tracking chambers providing
accurate muon tracking for $|\eta| < 2.7$ and fast detectors for triggering for $|\eta| < 2.4$.

Events are selected using a first-level trigger implemented in custom electronics,
which reduces the event rate to a design value of at most 100~kHz using a subset of detector information~\cite{Aaboud:2016leb}.
Software algorithms with access to the full detector information are then used in the high-level trigger to yield an average
recorded event rate of about 1~kHz.

%% file: sections/samples.tex
\section{Data and simulated event samples}
\label{sec:sample}

Data were collected in 2015 and 2016 using $pp$ collisions at a centre-of-mass energy of $\sqrt{s}=13$~\TeV\
with a bunch spacing of 25~ns.
The average number of $pp$ interactions per bunch crossing is 13 in 2015 and 25 in 2016, 
with a peak instantaneous luminosity up to \peakLumi.
Events from $pp$ collisions are recorded using a diphoton trigger with transverse energy ($\et$)
thresholds of 35~\GeV\ and 25~\GeV\
for the \et-ordered leading and subleading photon candidates, respectively. 
In the high-level trigger, the shapes of the energy depositions in the EM calorimeter are required to match those expected for electromagnetic showers initiated by photons.
The trigger has a signal efficiency close to 100\% for events fulfilling the final event selection, with an uncertainty below $0.4\%$.
After applying data-quality requirements,
the data sample corresponds to an integrated luminosity of \intLumiFifteen\ for the 2015 data and
\intLumiSixteen\ for the 2016 data.
The measurement of the integrated luminosity has an uncertainty of 2.1\% for the 2015 data and 3.4\% for the 2016 data.
The uncertainties in the 2015 and 2016 integrated luminosities are derived
following a methodology similar to that detailed in Ref.~\cite{Aaboud:2016hhf},
from a calibration of the luminosity scale using a $x$--$y$ beam-separation scan performed in August 2015, and a preliminary calibration using a scan performed in May 2016, respectively.
The correlation between the two years' luminosity uncertainties is taken into account.

Simulated Monte Carlo (MC) events are used for optimizing the search strategy~\cite{Aaboud:2016hhf},
and for the signal and background modelling studies detailed in Sections~\ref{sec:signal_modelling} and \ref{sec:background}, respectively.
Interference effects between the resonant signal and the background processes are neglected.

The spin-0 signal MC samples were generated using the effective-field-theory approach implemented in
\textsc{MadGraph5\_aMC@NLO}~\cite{mg5anlo} version 2.3.3 at next-to-leading order (NLO) in quantum chromodynamics (QCD).
From the Higgs characterization framework~\cite{HCframework}, 
CP-even dimension-five operators coupling the new resonance to gluons and photons were included.
Samples were generated with the NNPDF3.0 NLO parton distribution functions (PDFs)~\cite{nnpdf30nlo}, using the A14
set of tuned parameters (tune) of \textsc{Pythia 8.186}~\cite{Pythia8, ATL-PHYS-PUB-2014-021} for the parton-shower and hadronization simulation.
Simulated samples were produced for fixed values of the mass and width of the assumed resonance, spanning the range 200--2400~\GeV\
for the mass, and the range from 4~\MeV\ to 10\% of the mass for the decay width.
Choosing an improved signal model with an event generator different from the one used in Ref.~\cite{Aaboud:2016tru} provides a description of the signal which is less sensitive to modelling effects from the off-shell region.
The impact of this change is only visible in scenarios with a large signal decay width, with mass values at the \TeV\ scale.

Spin-2 signal samples for the RS1 model were generated using \textsc{Pythia 8.186},
with the NNPDF23LO PDF set~\cite{nnpdf23lo} and the A14 tune. 
Only the lightest KK graviton excitation was generated.
Its mass $m_{G^{*}}$ was varied in the range between 500~\GeV\ and 5000~\GeV.
The dimensionless coupling \coupling, where $\mplRed = \mpl/\sqrt{8\pi}$ is the reduced Planck scale
and $k$ the curvature scale of the extra dimension, is assumed to be in the range 0.01 to 0.3. 
For $\coupling < 0.3$, the KK graviton is expected to be a fairly narrow resonance~\cite{Davoudiasl:1999jd},
with a width given by $1.44 (\coupling)^2 m_{G^*}$.

The non-resonant KK graviton signal was simulated using the ADD model with the representation proposed by Giudice, Rattazzi and Wells (GRW)~\cite{Giudice:1998ck}.
The \textsc{Sherpa} event generator (version 2.1.1) was used to simulate both the SM and ADD processes at the same time, including their interference.
The ultraviolet cutoff scale \ms\ was varied between 3500~\GeV\ and 6000~\GeV\ in the simulation.

Events containing two prompt photons,
representing an irreducible background to the search, 
were simulated using the \textsc{Sherpa}~\cite{Gleisberg:2008ta} event generator, version 2.1.1. 
Matrix elements were calculated with up to two additional partons at leading order (LO) in QCD 
and merged with the \textsc{Sherpa} parton-shower simulation~\cite{Schumann:2007mg}
using the ME+PS@LO prescription~\cite{Hoeche:2009rj}.
The gluon-induced box process was also included.
The \textsc{CT10} PDF set~\cite{Lai:2010vv} was used in conjunction with a dedicated parton-shower tune of \textsc{Sherpa}. 
Samples of the photon+jet reducible background component were also generated using \textsc{Sherpa}, version 2.1.1,
with matrix elements calculated at LO with up to four additional partons.
The same PDF set, parton-shower tune and merging prescription as for the diphoton sample were used.
To study the dependence of data composition results (Section~\ref{sec:event_selection}) on the event generator,
\textsc{Pythia8} was also used to generate SM diphoton events, 
based on the LO quark--antiquark $t$-channel
diagram and the gluon-induced box process, and photon+jet events.
The same PDF set and parton-shower tune as for the RS1 model signal samples were used.

The generated events were passed through a full detector simulation~\cite{Aad:2010ah} based on \textsc{Geant4}~\cite{Geant4}. 
Pile-up from additional $pp$ collisions in the same or neighbouring bunch crossings was simulated by overlaying each MC
event with a variable number of simulated inelastic $pp$ collisions, 
generated using \textsc{Pythia8} with the AZNLO tune~\cite{AZNLO:2014}. The MC events were weighted to reproduce
the distribution of the average number of interactions per bunch crossing observed in the data.

%% file: sections/selection.tex
\section{Event selection}
\label{sec:event_selection}

The event selection criteria are the same as those described in Ref.~\cite{Aaboud:2016tru}.
Photon candidates are reconstructed from clusters of energy deposited in the EM calorimeter,
and may have tracks and conversion vertices reconstructed in the ID.
They are required to be in a fiducial region of the EM calorimeter defined by $|\eta| < 2.37$, not including the transition region $1.37<|\eta|<1.52$ between the barrel and end-cap calorimeters.
Compared to the processing used in Ref.~\cite{Aaboud:2016tru}, the reconstruction of converted photons from stand-alone TRT tracks was re-optimized to improve the efficiency and to cope with higher pile-up in the 2016 data-taking period.
For an average number of interactions per bunch crossing close to 40, the re-optimized algorithm maintains the same reconstruction efficiency for genuine photon conversions, while the rate of unconverted photons reconstructed as converted is reduced from around 20\% to around 5\%.
In addition, a small bias in the track parameters was corrected.
These changes lead to small event-by-event differences between the two reconstructions in the classification of the photons as either converted or unconverted.

The two photon candidates with the highest transverse energies in each event, satisfying $\et > 40$~\GeV\ and 30~\GeV, are retained. The energy measurement is based on a multivariate regression algorithm~\cite{Aad:2014nim} used to determine corrections to the energy of the clusters, developed and optimized on simulated events. The calibration of the energies deposited in each layer of the calorimeter, the overall energy scale and energy resolution are determined in situ. The regression algorithm was retrained to account for the small changes in the conversion reconstruction. For photons near the transition region between the barrel and end-cap calorimeters,
the information from the scintillators located in front of the end-cap cryostat usually improves the energy resolution by a few percent compared to Ref.~\cite{Aaboud:2016tru}, although in a few rare cases a significant shift of the measured energy may occur.
At \et\ values larger than 100--200~\GeV\ the energy resolution is dominated by the constant term 
of the calorimeter energy resolution, which varies from 0.9\% to 2.0\% in different $\eta$ regions.
The uncertainty in the photon energy scale for $\et > 100$~\GeV\ is typically 0.5--1.5\% depending on $\eta$.
The uncertainty in the photon energy resolution is driven by the uncertainty in the constant term in the \et\ range relevant to this analysis.
At $\et = 300~\GeV$, the relative uncertainty is 30--40\% depending on $\eta$.

Photons are required to fulfil \emph{tight} identification criteria~\cite{Aaboud:2016yuq} based on variables that measure the shape of the electromagnetic showers in the calorimeter (``shower shapes''), in particular in the finely segmented first layer. The efficiency of the photon identification increases with \et\ from 90\% at 50~\GeV\ to 95\% at 200~\GeV. The associated $\eta$-dependent uncertainties were measured in the whole dataset from 2015 and 2016 using the same methods as in Ref.~\cite{Aaboud:2016yuq}. They vary between 0.2\% and 4\% below 200~\GeV, and between 1\% and 4\% above.

To further reject the background from jets misidentified as photons, the candidates are required to be isolated using both calorimeter and tracking detector information. The scalar sum of the \et\ of energy clusters within a cone of size $\Delta R = \sqrt{(\Delta\eta)^2 + (\Delta\phi)^2} = 0.4$ around the photon candidate, excluding the photon energy deposits and correcting for pile-up and underlying-event contributions~\cite{STDM-2010-08,Cacciari:2008gn,Cacciari:2009dp}, is required to be below $0.022 \et + 2.45$~\GeV,
where $\et$ is the transverse energy of the photon candidate. The sum of the tranverse momentum (\pt) of tracks within $\Delta R = 0.2$ of the photon candidate, not including tracks associated with a photon conversion, is required to be below $0.05 \et$.
In order to minimize pile-up effects and to improve the separation between signal and background processes, tracks with $\pt < 1$~\GeV\ are excluded from the sum. Tracks are also excluded if they have a large impact parameter with respect to the primary vertex identified by combining the photon direction measurements from the calorimeter with the tracking information from the ID as described in Ref.~\cite{Aaboud:2016tru}.
The efficiency of the combined isolation requirement for photons fulfilling the tight identification selection in signal MC
samples is 88\% to 97\% in the \et\ range from 100~\GeV\ to 500~\GeV, with an uncertainty around 1\%. Compared to Ref.~\cite{Aaboud:2016tru}, improvements in the selection of tracks associated with photon conversions lead to a few percent increase in the  efficiency of the isolation requirement for genuinely converted photons.

Different kinematic selections~\cite{Aaboud:2016tru} are applied in the searches for spin-0 and spin-2 signals to exploit the kinematic properties of the decay photons. In the selection used to search for a spin-0 resonance, the transverse energy is required to be $\et > 0.4 m_{\gamma\gamma}$ for the leading photon 
and $\et > 0.3 m_{\gamma\gamma}$ for the subleading photon, for a given value of the diphoton invariant mass $m_{\gamma\gamma}$.
With these requirements, \yieldSpinZero\ (\yieldSpinZeroHighMass)~events are selected in the data with $m_{\gamma\gamma} >150$~\GeV\ ($>200$~\GeV). In the selection used to search for spin-2 resonant and non-resonant signals, the transverse energy of each photon is required to be \et~$>$~55~\GeV.
With these requirements, \yieldSpinTwo\ diphoton events with $m_{\gamma\gamma} >$~200~\GeV\ are selected in the data. 
The search for a non-resonant signal uses only events with $\mgg>2240$~\GeV,
which is optimized for the expected limit on the ultraviolet cutoff scale \ms\ in the ADD model using the signal and background samples described in Section~\ref{sec:sample}.

\begin{figure}[!]
\begin{center}
\subfloat[]{\includegraphics[width=.5\textwidth]{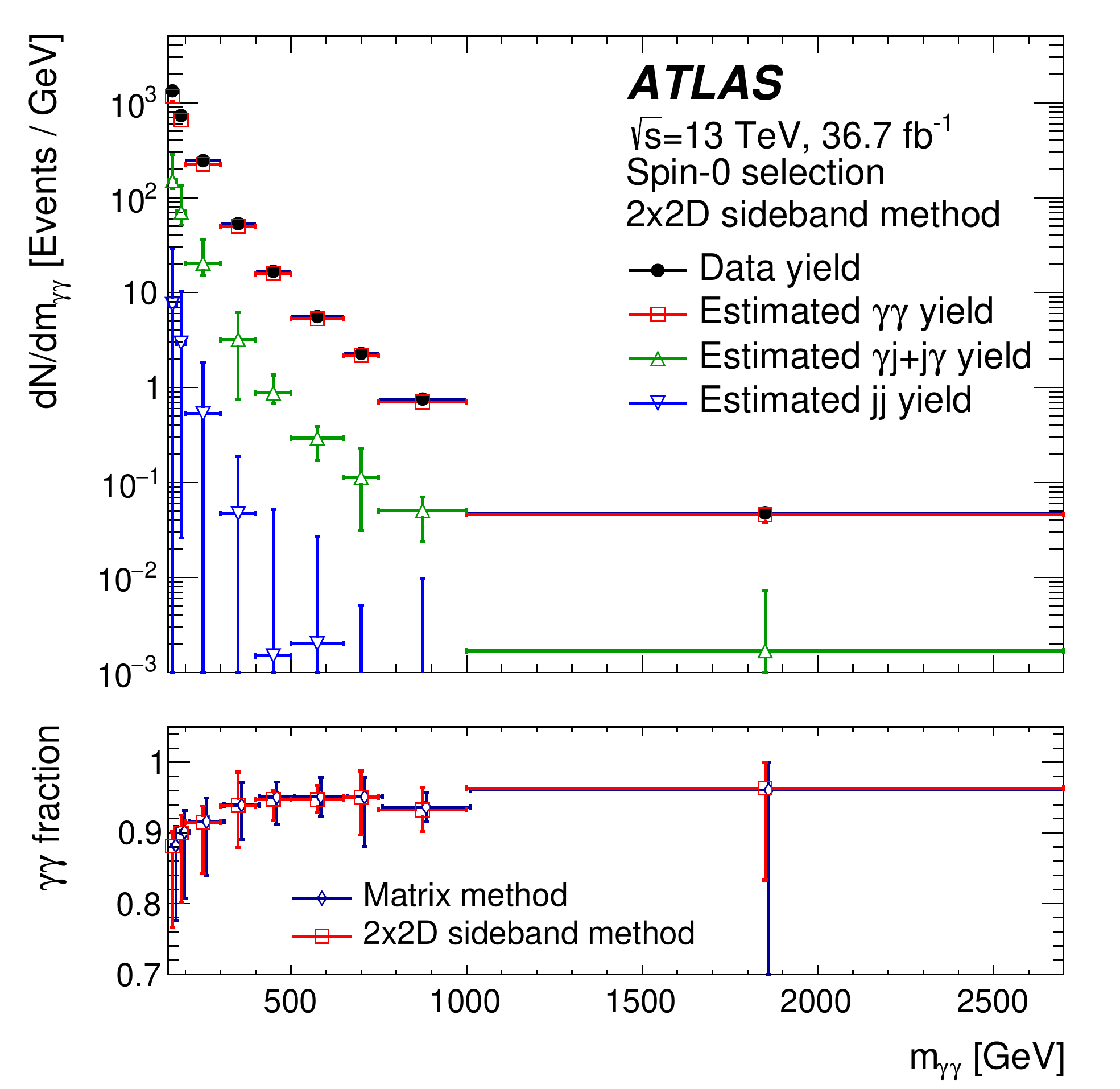}}
\subfloat[]{\includegraphics[width=.5\textwidth]{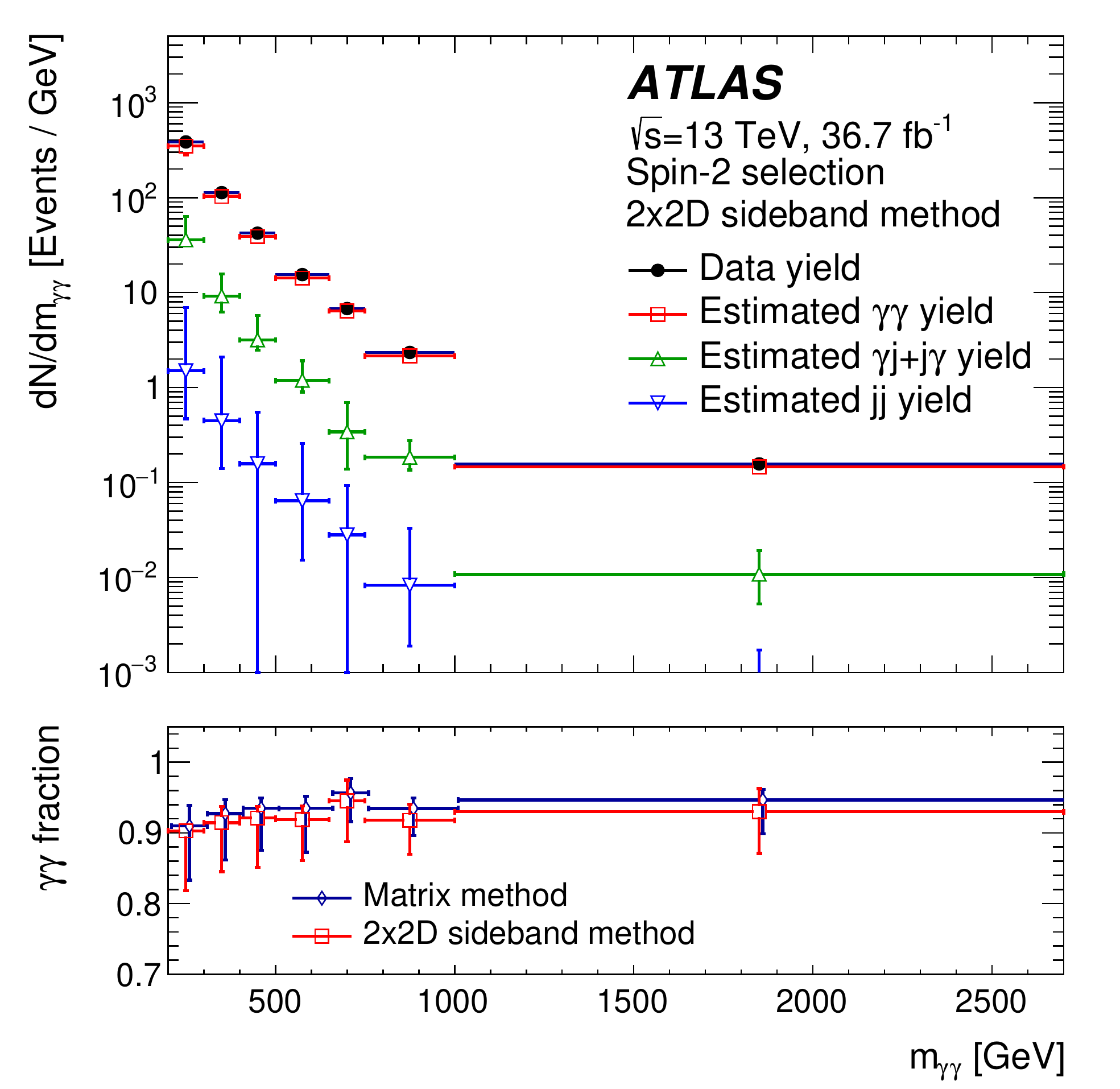}}
\end{center}
\caption{The diphoton invariant-mass distributions of the data are shown in the upper panels for (a) the spin-0 
and (b) the spin-2 selections and their decomposition into contributions 
from genuine diphoton ($\gamma\gamma$), photon+jet ($\gamma j$ and $j \gamma$) and dijet ($jj$) events as determined using the 2x2D sideband method.
The bottom panels show the purity of diphoton events as determined by the matrix method and the 2$\times$2D sideband method.
Each point in the distributions is plotted in the centre of the corresponding bin.
The total uncertainties, including statistical and systematic components added in quadrature, are shown as error bars.
}
\label{fig:bkg_composition}
\end{figure}

The composition of the selected data sample is studied using three methods detailed in Ref.~\cite{Aaboud:2016tru}.
The 2$\times$2D sideband method and the matrix method are based on sidebands constructed by inverting photon identification and isolation requirements.
The results from these two methods are in good agreement as shown in Figure~\ref{fig:bkg_composition}.
The selected samples consist mainly of events from diphoton production, with a purity estimated by the 2$\times$2D sideband method to be $(91^{+3}_{-7})$\% for the spin-0 selection and $(91^{+3}_{-8})$\% for the spin-2 selection,
increasing by a few percent with \mgg.
Uncertainties in these purity estimates originate from the statistical uncertainty in the data sample,
the definition of the control region failing the tight identification requirement,
the event generator dependence (difference between \textsc{Sherpa 2.2.1} and \textsc{Pythia8}),
the modelling of the isolation and shower-shape distributions,
and possible correlations between the isolation variables and the inverted identification criteria.
The remaining background is mostly composed of photon+jet and dijet production, with one or two jets misidentified as photons. Backgrounds from other sources are negligible.
The composition derived from the 2$\times$2D sideband method is used to select the function to model the background in the spin-0 resonance search.
In the background estimate used for the spin-2 resonance and non-resonant signal searches, the composition of the data
sample is determined by a third method, which exploits the isolation profile of the two photons in the calorimeter, with consistent results obtained for the spin-2 selection as mentioned above.

%% file: sections/signal_modeling.tex
\section{Signal modelling}
\label{sec:signal_modelling}

For both spin-0 and spin-2 resonance searches, parametric models of the diphoton invariant-mass distributions are used in order to test for different resonance masses and widths or \coupling. The distribution for a signal of given mass and width is obtained by convolving the detector resolution with the predicted mass line-shape distribution at the particle level.
The detector resolution is modelled by a double-sided Crystal Ball (DSCB) function, composed of a Gaussian core with power-law tails~\cite{Aaboud:2016tru}.
The line-shape at the particle level for each signal model is taken to be the product of the analytic differential cross-section expression for particle production and decay to photons, and a parametrised form of the parton luminosity.
The signal model is in good agreement with invariant-mass distributions from simulated signal samples introduced in Section~\ref{sec:sample} with corresponding values of resonance mass and width (\coupling) for the spin-0 (spin-2) case.
Potential differences in the tails between the signal model and the simulation are found to have negligible impact on the extracted signal yield.
For relative widths comparable to or below the detector resolution of 1\% ($\coupling \lesssim 0.08$), the searches have limited sensitivity to the decay width.

In the spin-0 resonance search, a fiducial region at particle level that closely follows the selection criteria applied to the reconstructed data is defined for setting limits:
$|\eta| < 2.37$ and $\et > 0.4\mgg$ ($0.3\mgg$) for the leading (subleading) photon. In addition, an isolation requirement of $\ETiso < 0.05\et + 6~\GeV$ is applied to reproduce the selection applied at the reconstruction level. 
The particle-level isolation is calculated using all particles with lifetime greater than 10~ps at the event-generator level in
a cone of $\Delta R = 0.4$ around the photon direction.
Compared to the fiducial region definition used in Ref.~\cite{Aaboud:2016tru},
the mass range requirement introduced to reduce the model-dependence from the off-shell region is removed since its impact is found to be negligible using the new \textsc{MadGraph5\_aMC@NLO} signal model.
The combined reconstruction and identification efficiency, defined as the ratio of the number of events fulfilling all the selections placed on reconstructed quantities to the number of events in the fiducial acceptance, varies from 64\% at 200~\GeV\ to 75\% at 2700~\GeV.
It is evaluated with signal MC samples corresponding to the narrow-width approximation (NWA, $\Gamma_X = 4~\MeV$), with a 2.8\% uncertainty assigned to cover variations in the decay width (0--10\%).

In the spin-2 resonance search, the results are evaluated assuming the acceptance as well as the reconstruction and identification efficiencies obtained by the MC simulation of KK graviton decays.
The product of these two terms is evaluated as a function of the KK graviton mass \mG\ using MC samples with $\coupling=0.2$.
It increases from 45\% at 500~\GeV\ to 65\% at 5000~\GeV, with a 2.9\% uncertainty that results from varying \coupling.

In the non-resonant signal search, 
the acceptance and reconstruction and identification efficiencies are evaluated for the excess yield in the presence of an ADD signal, which is defined as the difference between the sum of ADD and SM contributions (including their interference) and the SM contribution.
The acceptance for the excess yields in the signal region increases from 58\% at $\ms=3.5~\TeV$ to 65\% at $\ms=5~\TeV$. For larger values of \ms, the acceptance decreases to about 58\% at 8~\TeV\ due to a larger contribution from the interference term, which has smaller acceptance. The combined reconstruction and identification efficiency for the excess yields at different $\ms$ values is approximately constant at 77\% within MC statistical uncertainties.

Uncertainties in the signal parameterization and in the acceptance and detector efficiency correction factors for the signal considered in each search are summarized in Table~\ref{tab:uncertainties_signal}.

\begin{table}
\centering
{\small
\begin{tabular}{lccc}
\hline
\multirow{2}{*}{Uncertainty source}           &    Spin-0 resonance & Spin-2 resonance & Spin-2 non-resonant \\
           &    [\%] & [\%] & [\%] \\
\hline
Signal mass resolution  &  17--38 & 28--36  & -- \\
\hline
Signal photon identification efficiency  &  1.3--3.0 & 2.6--3.1  & 3.2 \\
Signal photon isolation efficiency   &  1.1--1.3 & 1.2--1.4   &  1.4 \\
Signal width dependence &     2.8  & 2.9      &    --   \\
Trigger efficiency                      & 0.4 & 0.4 & 0.4     \\
Luminosity               &   \multicolumn{3}{c}{2.1 (2015), 3.4 (2016)}  \\
\hline
Total uncertainty in signal yield & 4.6--5.4 & 5.3--5.5 & 4.8 \\
\hline
\end{tabular}
}
\caption{Summary of the relative systematic uncertainties (in percent). For mass-dependent uncertainties, the quoted values cover the range from 200~\GeV\ (500~\GeV) to 2700~\GeV\ (5000~\GeV)
  for the spin-0 (spin-2) resonance search. The uncertainty in the total signal yield corresponds to the sum in quadrature of the individual components, not including the uncertainty in the mass resolution.
}
\label{tab:uncertainties_signal}
\end{table}

%% file: sections/background_estimates.tex
\section{Background estimates}
\label{sec:background}
Two different methods~\cite{Aaboud:2016tru} are used to estimate the SM background contributions to the $m_{\gamma\gamma}$ distribution. The approach adopted in the spin-0 search, appropriate for a mass range with enough data close to the investigated resonance mass,
is based on a fit using a smooth functional form,
with parameter values determined simultaneously with the signal and background yields by the fit. The mass distribution is fitted in the range above 180~\GeV\ (or 150~\GeV\ when fitting 2015 data alone),
and the search range for the signal is 200--2700~\GeV.
The procedure detailed in Ref.~\cite{HIGG-2013-08} is used to check that the functional form is flexible enough to accommodate different physics-motivated underlying background distributions from MC simulations, within the uncertainties in the measured background composition and PDF set. The potential bias due to the choice of the functional form is estimated by the fitted signal yield (``spurious signal'') in these background distributions,
and is considered as a systematic uncertainty.
The spurious signal is required to be less than 30\% of the statistical uncertainty in the fitted signal yield (from the background distributions) over most of the investigated mass range.
Below 400~\GeV, where the statistical uncertainty in the MC sample used to determine the spurious signal uncertainty is comparable to the maximum spurious signal allowed, the criterion is relaxed to 50\%.

In the spin-2 resonance search and also in the non-resonant signal search,
both of which target KK graviton signals at the \TeV\ scale, 
the small number of data events at high $m_{\gamma\gamma}$ values does not effectively constrain the invariant-mass distribution of the background.
The shape of the distribution of the irreducible diphoton background is thus predicted using the \textsc{Diphox}~\cite{Binoth:1999qq} computation at NLO in QCD,
which is found to be in good agreement with the predictions from \textsc{Sherpa} version 2.2.2~\cite{Siegert:2016bre} using the same order in QCD.
The background from photon+jet and dijet events is added using control samples from the data with the same method as discussed in Ref.~\cite{Aaboud:2016tru}.
An alternative method, based on the rate of jets misidentified as photons extracted from inclusive photon data and applied to similar control samples,
gives compatible results.
The different background components are combined according to the decomposition studies in data reported in Section~\ref{sec:event_selection} for the $m_{\gamma\gamma}$ distribution of the total background.
The normalization of the background is a free parameter in the maximum-likelihood fit to the data spectrum.
Uncertainty in the total background's shape results from uncertainties in both the shape and the relative normalization of each component~\cite{Aaboud:2016tru},
including the shape of the reducible background,
the relative normalization of the reducible and irreducible backgrounds,
the impact of the parton-level isolation requirement in \textsc{Diphox},
and the effect of the uncertainties in the scales and PDF set used in the \textsc{Diphox} computation. 
These uncertainties are taken to be correlated across the entire mass range. Statistical uncertainties in the simulated samples are also considered in each bin.
The uncertainty from the parton-level isolation requirement in \textsc{Diphox} is constrained by the low-mass sideband of the data \mgg\ spectrum in the maximum-likelihood fit, and thus reduces to about 1\% across most of the mass range.

In the non-resonant signal search, the SM yield and the associated uncertainties are computed directly as integrals of the above background estimates in the signal region $\mgg>2240$~\GeV, before the likelihood maximization.

The systematic uncertainties in the description of the background shapes are summarized  in Table~\ref{tab:uncertainties_background}.

\begin{table}
\begin{center}
{\small
  \begin{tabular}{lcc}
    \hline
    Uncertainty source & \multicolumn{2}{c}{Spin-0 resonance}\\
    \hline
    \multirow{2}{*}{Spurious signal [events]} & Narrow width & 74--0.006\\
    & 10\% relative width & 195--0.04\\
    \hline
    \hline
    Uncertainty source & Spin-2 resonance [\%] & Spin-2 non-resonant [\%] \\
    \hline
    Scales and PDFs in \textsc{Diphox} computation & 1--19 & 20 \tspp \\[3mm]
    Shape of the reducible background                   & 1--10 & 11 \tspp \\[3mm]
    Relative normalization of              & \multirow{2}{*}{1--2\, } & \multirow{2}{*}{2} \\
    reducible and irreducible backgrounds  &  & \\[3mm]
    Parton-level isolation requirement    & \multirow{2}{*}{10--12\, } & \multirow{2}{*}{9} \\
    in \textsc{Diphox}                    & &  \\[3mm]
    MC statistical uncertainty         & \multicolumn{2}{c}{$<1$} \\
    \hline
    Total & 10--25 & 25 \\
    \hline
    
  \end{tabular}
  }
\end{center}
\caption{Summary of pre-fit relative systematic uncertainties in the background estimation used in the searches.
  For mass-dependent uncertainties, the quoted mass ranges cover 200~\GeV\ (500~\GeV) until 2700~\GeV\
  for the spin-0 (spin-2) resonance search.
  In the spin-2 searches the PDF uncertainty in the irreducible background component dominates the total uncertainty beyond 2700~\GeV, rising to 130\% at 5000~\GeV.}
\label{tab:uncertainties_background}
\end{table}

%% file: sections/statistical_procedure.tex
\section{Statistical procedure}
\label{sec:stat}
In the spin-0 and spin-2 resonance searches, the numbers of signal and background events are estimated from maximum-likelihood fits of the signal-plus-background models to the corresponding
$m_{\gamma\gamma}$ distribution of the selected events.
In the search for an ADD graviton signal,
a counting experiment is performed in the region $\mgg>2240$~\GeV\
with the excess and the SM yields extracted using the methods discussed in Sections~\ref{sec:signal_modelling} and \ref{sec:background}, respectively.
Systematic uncertainties summarized in Tables~\ref{tab:uncertainties_signal} and \ref{tab:uncertainties_background}
are included in the fits via nuisance parameters constrained by Gaussian or log-normal penalty terms.

The $p$-value is determined from a profile-likelihood-ratio-test statistic~\cite{Cowan:2010js} as detailed in Ref.~\cite{Aaboud:2016tru}.
For the resonance searches, the local $p$-value for compatibility with the background-only hypothesis when testing a given
signal hypothesis ($p_0$) is evaluated based on the asymptotic approximation~\cite{Cowan:2010js}, and expressed in standard deviations in the following.
Global significance values are computed from background-only pseudo-experiments to account for the trial factors due to scanning both the signal mass and the width hypotheses.
The expected and observed 95\% confidence level (CL) exclusion limits
on the cross section times branching ratio to two photons are computed using 
a modified frequentist approach CL$_{\text{s}}$~\cite{cls} with
the asymptotic approximation to the test-statistic distribution~\cite{Cowan:2010js} for all three searches.
For resonance searches, cross-checks with sampling distributions generated using pseudo-experiments are performed for a few signal mass points
across the searched mass range. Below 2500~\GeV, the difference between the observed and expected limits is at the 0.01~fb level, which is well covered by the $\pm 1\sigma$ limit band.
In the mass range above 2500~\GeV\ covered by the spin-2 resonance search, the asymptotic approximation is no longer valid due to the small number of data events.
The observed and expected limits in this region are hence determined with pseudo-experiments.

%% file: sections/results.tex
\section{Results}
\label{sec:results}

Since the results of the 2015 data analysis were published in Ref.~\cite{Aaboud:2016tru}, photon reconstruction and energy calibration have been improved. The results for both the spin-0 and spin-2 resonance searches are updated as summarized below. In the light of the modest excess observed in the 2015 data alone, results are reported first considering the 2015 and 2016 data samples individually, before reporting the combined results.

\begin{itemize}
\item \textbf{Spin-0}:
  the largest local deviation from the background-only hypothesis in the 2015 dataset
  is \localSigSpinZeroFifteen\,$\sigma$ at a mass of \bestMassSpinZeroFifteen~\GeV\ with a relative width of \bestRelativeWidthSpinZeroFifteen.
  The results obtained on the same dataset with the old reconstruction and calibration, published in Ref.~\cite{Aaboud:2016tru}, presented a deviation of 3.9\,$\sigma$ at 750~\GeV\ with a relative width of 6\%.
\item \textbf{Spin-2}:
  the largest local deviation from the background-only hypothesis in the mass range 700--800~\GeV\ in the 2015 dataset is \localSigSpinTwoFifteen\,$\sigma$ at a mass of \bestMassSpinTwoFifteen~\GeV\ with a \coupling\ value of \bestCouplingSpinTwoFifteen.
  The results obtained on the same dataset with the old reconstruction and calibration, published in Ref.~\cite{Aaboud:2016tru}, presented a deviation of 3.8\,$\sigma$ at 750~\GeV\ with a \coupling\ value of 0.23.
\end{itemize}

In the 2016 dataset, the observations are summarized as follows.
\begin{itemize}
\item \textbf{Spin-0}: 
  the largest local deviation from the background-only hypothesis corresponds to a 2.0\,$\sigma$ narrow-width excess at 304~\GeV.
  Within the mass interval 700--800~\GeV, the largest local deviation from the background-only hypothesis corresponds to a \localSigSpinZeroSixteen\,$\sigma$ \bestWidthSpinZeroSixteen\ excess at \bestMassSpinZeroSixteen~\GeV.
\item \textbf{Spin-2}: the largest local deviation from the background-only hypothesis 
  corresponds to a \localSigSpinTwoSixteen\,$\sigma$ excess at \bestMassSpinTwoSixteen~\GeV\ with a best-fit \coupling\ value of \bestCouplingSpinTwoSixteen.
\end{itemize}

The results combining the 2015 and 2016 datasets are summarized below. A complete set of tables with the full limit results, including those from additional width and \coupling\ scenarios not covered in this Letter, are available at the Durham HepData repository.

\begin{figure}[t]
\begin{center}
  \subfloat[\label{fig:mgg_distributions_spin0}]{\includegraphics[width=.49\textwidth]{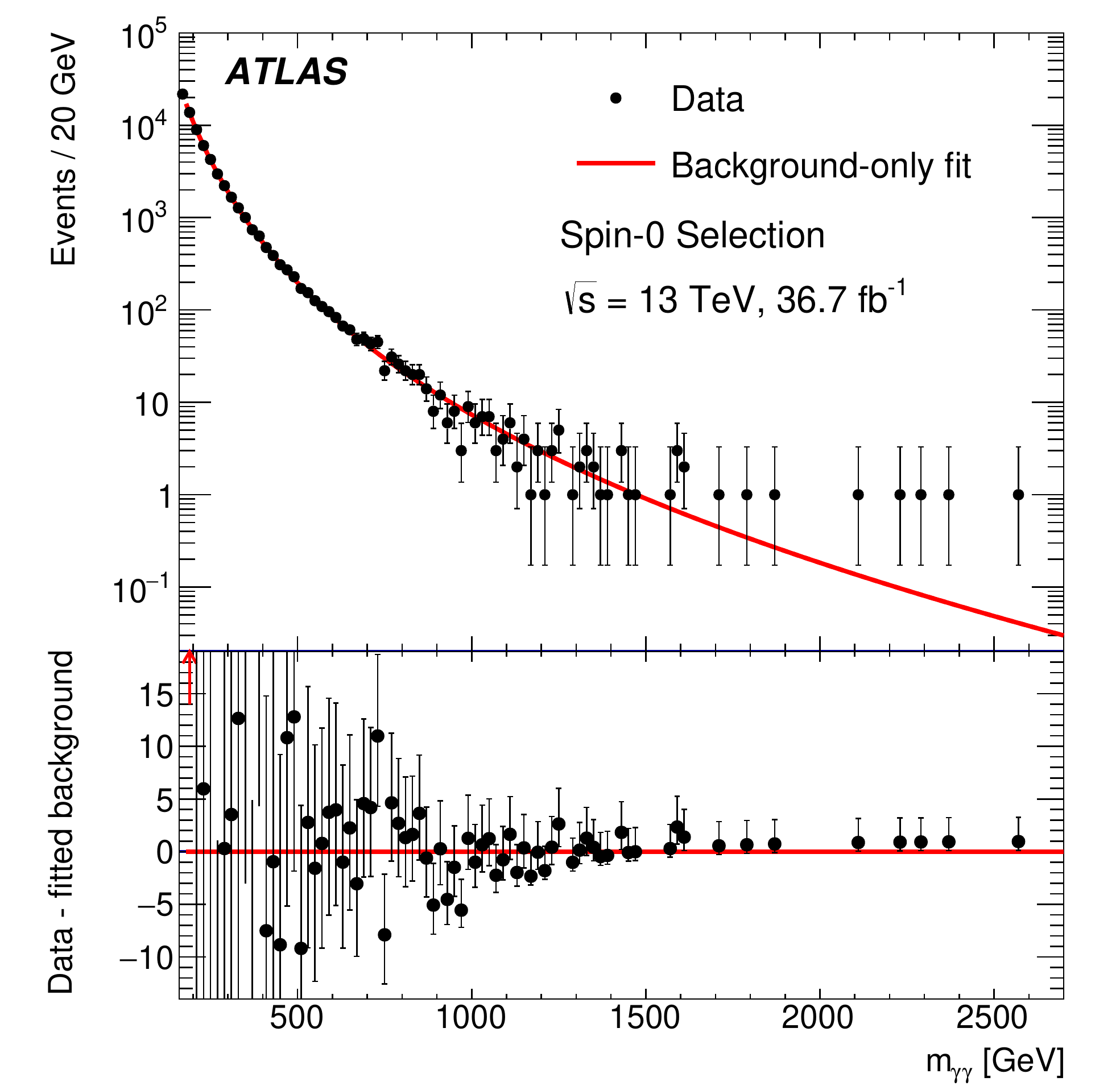}}
  \subfloat[\label{fig:mgg_distributions_spin2}]{\includegraphics[width=.49\textwidth]{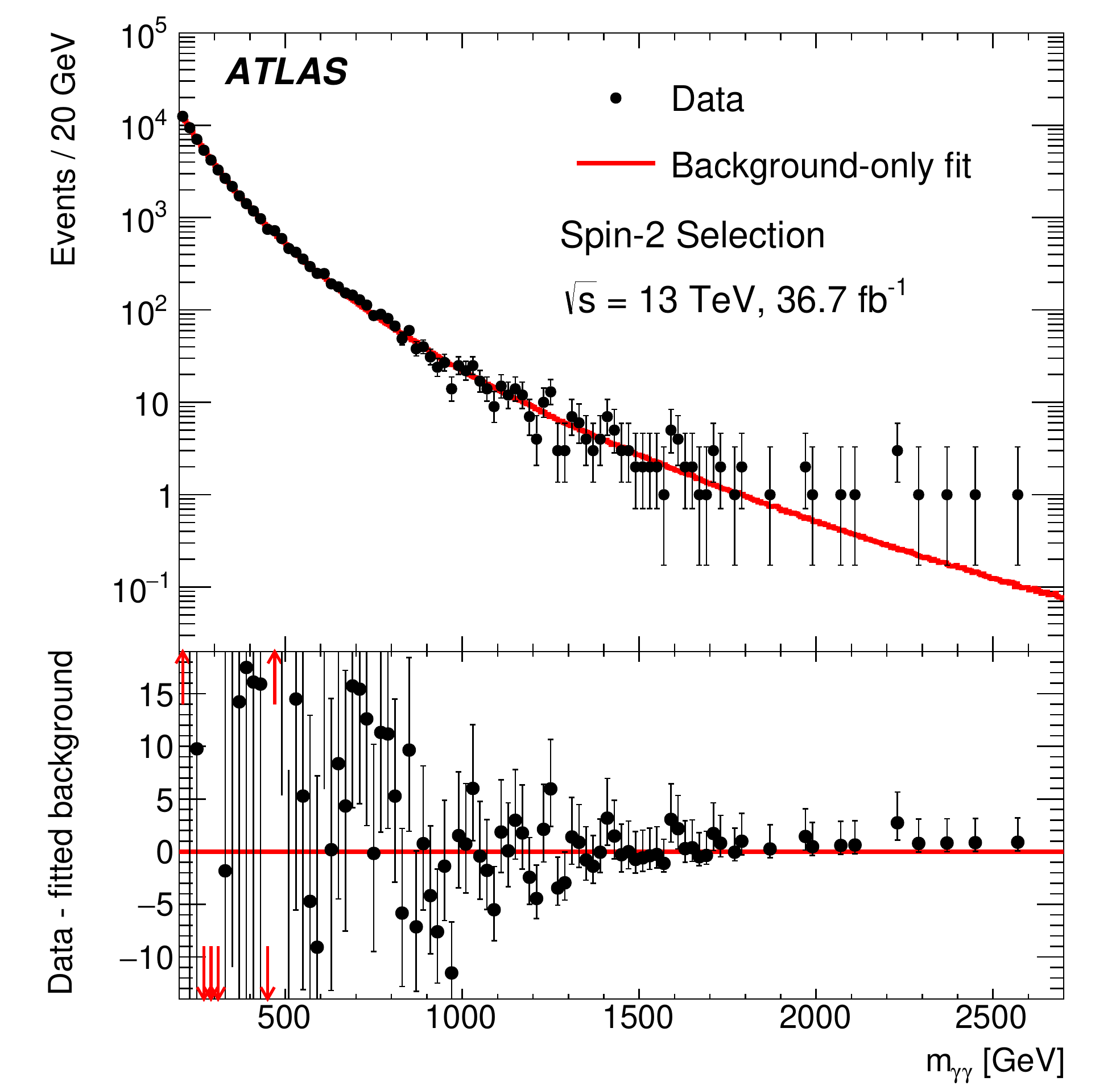}}  
\end{center}
\caption{Distributions of the diphoton invariant mass for events passing (a) the spin-0 selection or (b) the spin-2 selection,
  with the background-only fits superimposed.
  The data points are plotted at the centre of each bin.
  The error bars indicate statistical uncertainties.
  The differences between the data and the fits are shown in the bottom panels.
  The arrows in the lower panels indicate values outside the range by more than one standard deviation.
  There is no data event with $m_{\gamma\gamma}>2700$~\GeV.
}
\label{fig:mgg_distributions}
\end{figure}

\begin{figure}[t]
  \begin{center}
  \subfloat[\label{fig:p0_2D_spin0}]{\includegraphics[width=.49\textwidth]{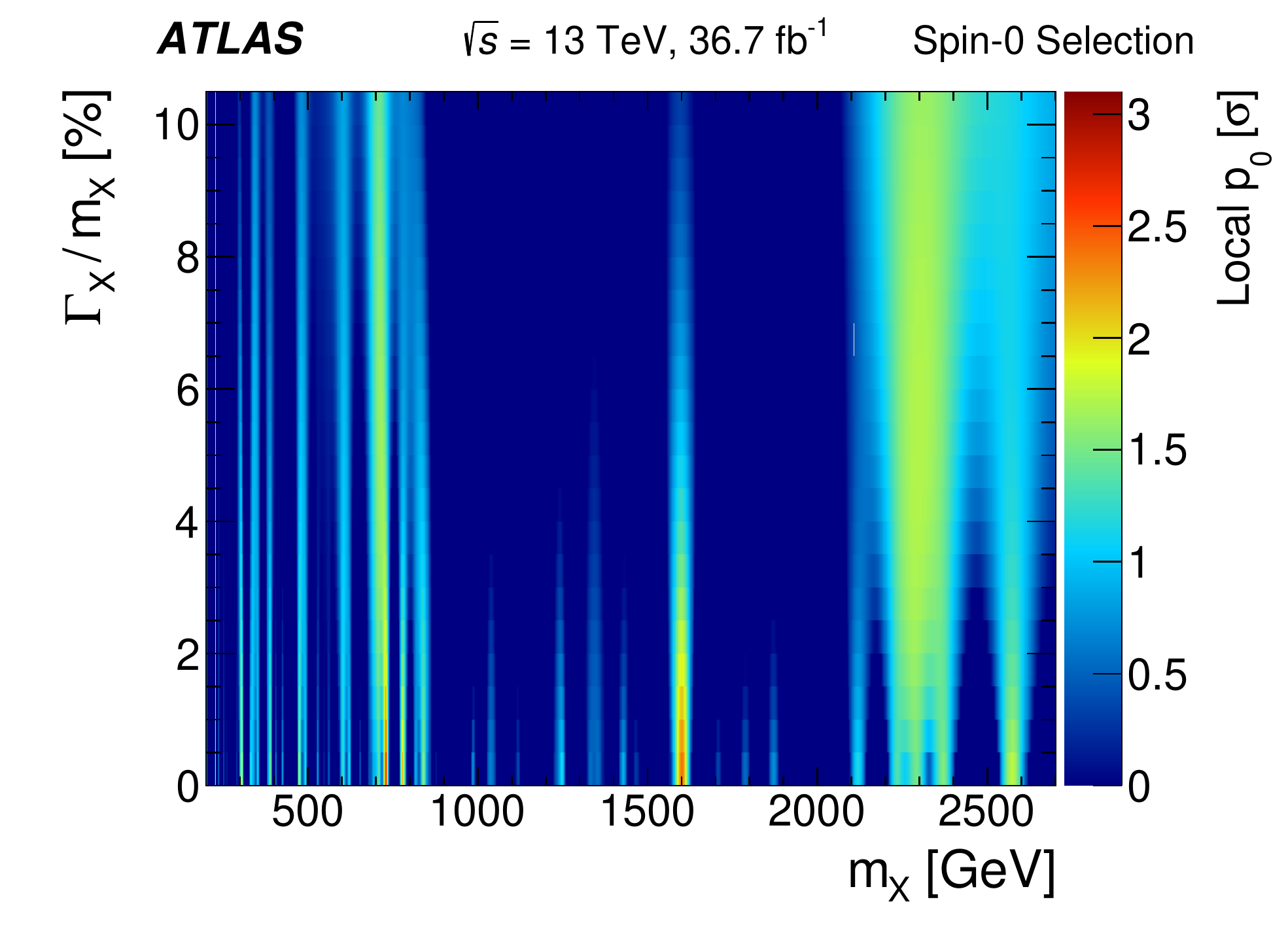}}
  \subfloat[\label{fig:p0_2D_spin2}]{\includegraphics[width=.49\textwidth]{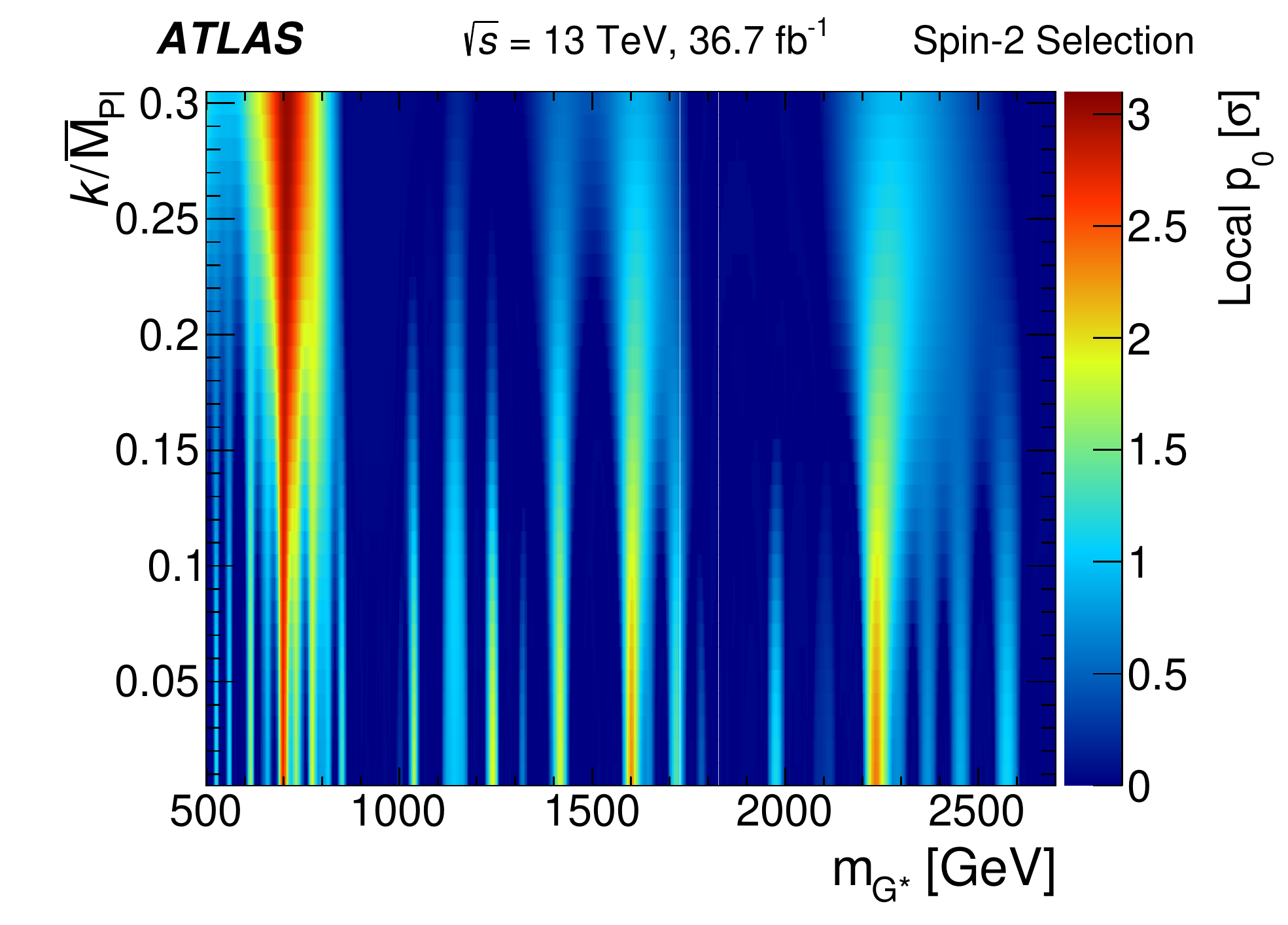}}
      \end{center}
  \caption{Compatibility, in terms of local $p_0$ quantified in standard deviations $\sigma$,
    with the background-only hypothesis (a) as a function of the assumed signal mass $m_X$ and relative width $\Gamma_X/m_X$ for the spin-0 resonance search and (b) as a function of the assumed signal mass $m_{G^*}$ and \coupling\ for the spin-2 resonance search. Only positive excesses are considered.
  }
  \label{fig:p0_2D}
\end{figure}

\begin{figure}[!ht]
\begin{center}
 \subfloat[\label{fig:limit_spin0}]{\includegraphics[width=.49\textwidth]{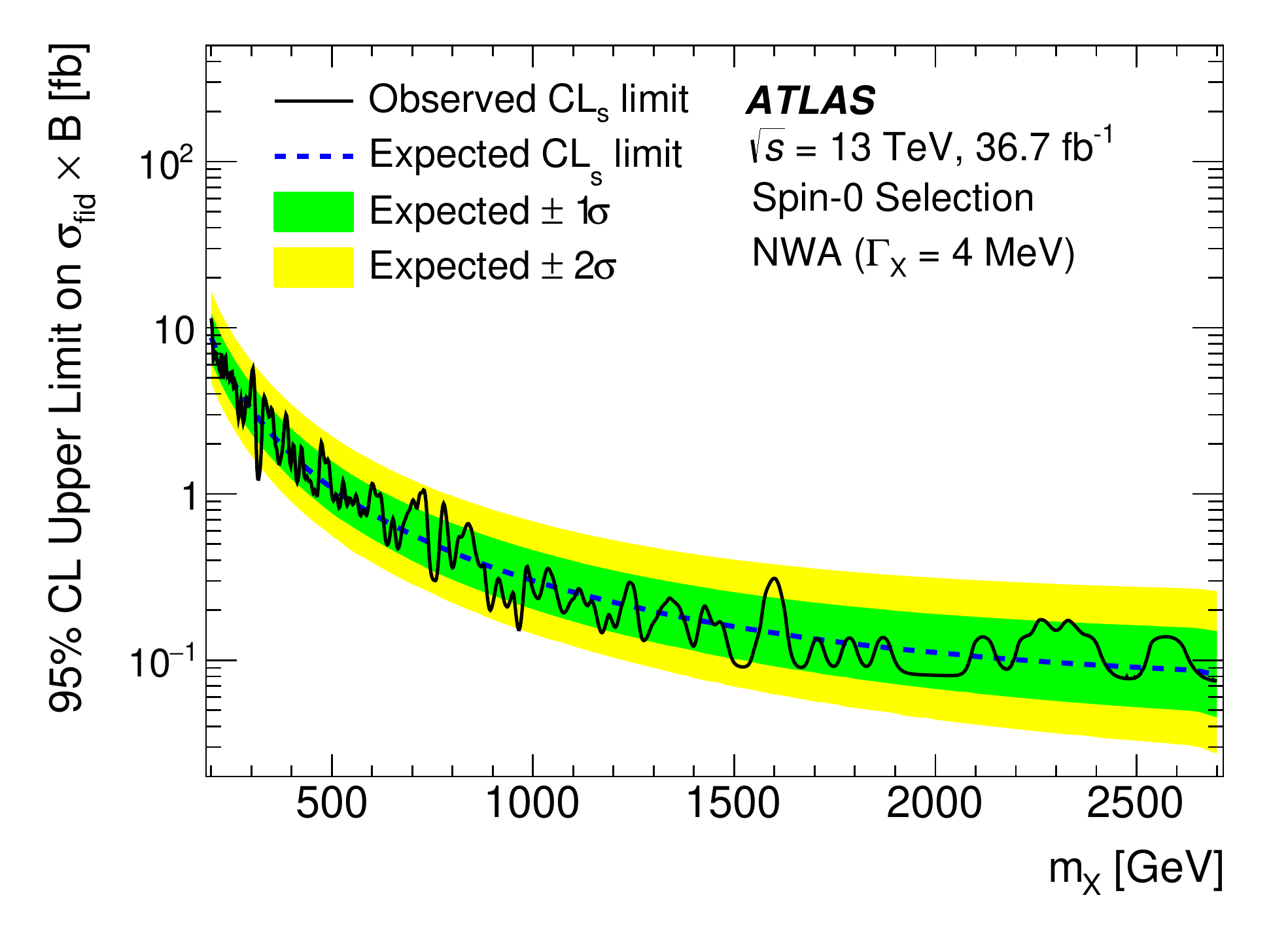}}
 \subfloat[\label{fig:limit_spin2}]{\includegraphics[width=.49\textwidth]{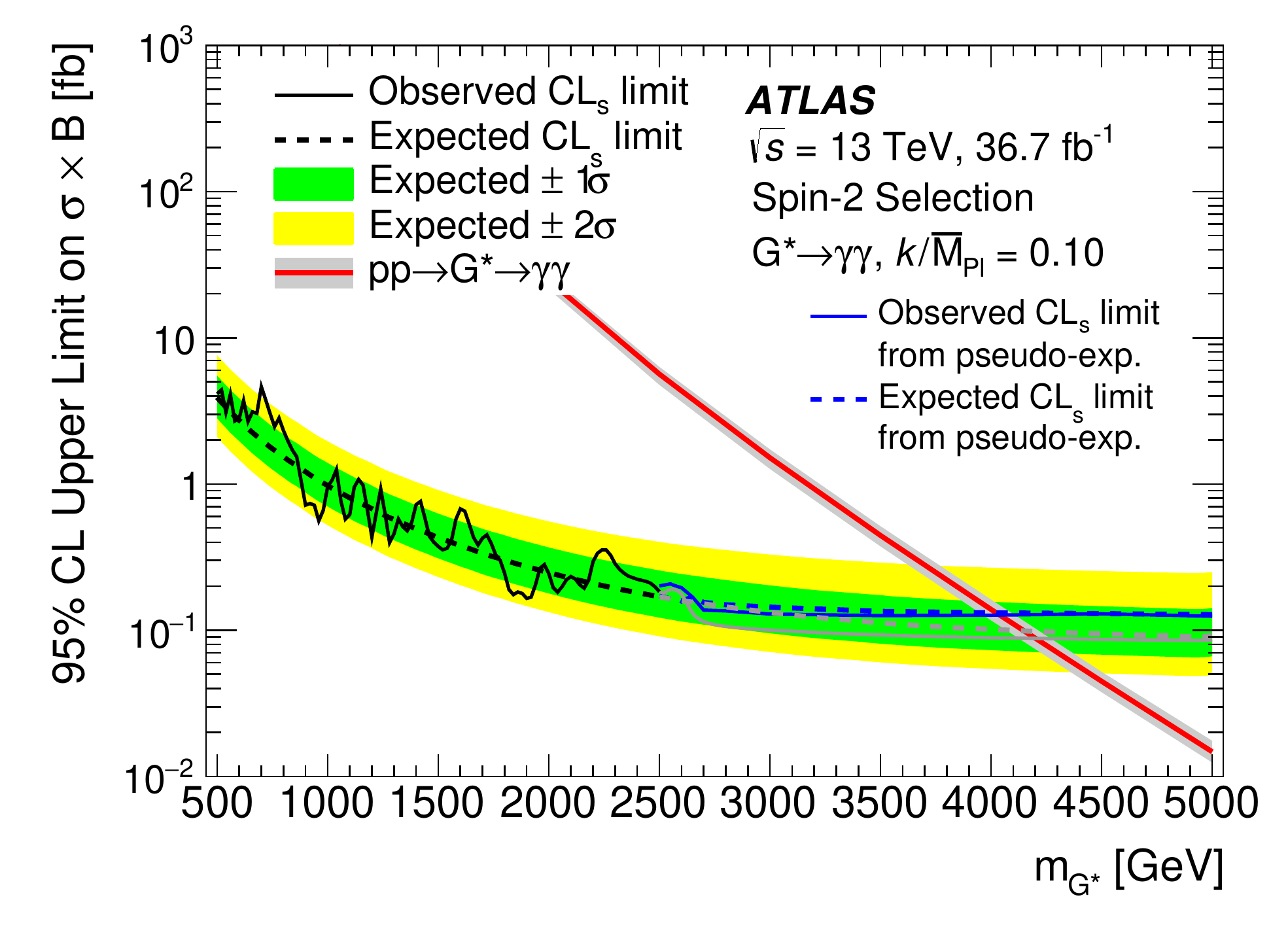}}
\end{center}
\caption{(a) Upper limits on the fiducial cross section times branching ratio to two photons at $\sqrt{s} = 13$~\TeV\ of a narrow-width ($\Gamma_X$ = 4~\MeV) spin-0 resonance as a function of its mass $m_X$. (b) Upper limits on the production cross section times branching ratio to two photons at $\sqrt{s} = 13$~\TeV\ of the lightest KK graviton as a function of its mass $m_{G^*}$ for $\coupling=0.1$.
  For $m_{G^*}>2500~\GeV$, the observed and expected limits are determined with pseudo-experiments shown by the blue solid and dashed lines, respectively.
 Predictions are shown for the RS1 model, where the grey shaded band represents the PDF uncertainty.
}
\label{fig:limit}
\end{figure}

\begin{itemize}
\item \textbf{Spin-0}:
  the diphoton invariant-mass distribution of the events passing the spin-0 selection is shown in Figure~\ref{fig:mgg_distributions_spin0}.
  The compatibility of the data with the background-only hypothesis as a function of both the assumed mass and width values is
  shown in Figure~\ref{fig:p0_2D_spin0}.
  The largest deviation from the background-only hypothesis is observed at a mass of \bestMassSpinZero~\GeV\ for a narrow width,
  with a local $p_0$ of \localSigSpinZero\,$\sigma$.
  The corresponding global significance is null,
  as the local deviation is less than the median largest deviation in background-only pseudo-experiments in the search region defined by 200--2700~\GeV\ in resonance mass and 0--10\% in relative width.
  Figure~\ref{fig:limit_spin0} shows the upper limits on the signal fiducial cross section times branching ratio to two photons for a narrow-width (4~\MeV) spin-0 resonance as a function of its mass.
  The limit on the fiducial cross section times branching ratio ranges from \maximumLimitSpinZeroNWA\ 
  at 200~\GeV\ to about \minimumLimitSpinZeroNWA\ at 2700~\GeV.
  The impact of the systematic uncertainties on the expected limit decreases with the resonance mass from 29\% at 200~\GeV\ to 5\% at 700~\GeV. Above 700~\GeV, the impact is typically 2--3\%.
\item \textbf{Spin-2}:
  the invariant-mass distribution of the events passing the spin-2 selection is shown in Figure~\ref{fig:mgg_distributions_spin2}.
  The local compatibility of the data with the background-only hypothesis
  as a function of both the resonance mass and \coupling\ values
  is shown in Figure~\ref{fig:p0_2D_spin2}.
  The largest deviation from the background-only hypothesis in the combined dataset is observed for a mass of \bestMassSpinTwo~\GeV, and $\coupling$ of \bestCouplingSpinTwo,
  corresponding to \localSigSpinTwo\,$\sigma$ local $p_0$.
  The global significance, estimated from pseudo-experiments in the search region of 500--2700~\GeV\ in mass and 0.01--0.3 in \coupling,
  is \globalSigSpinTwo\,$\sigma$.
  Figure~\ref{fig:limit_spin2} shows the limits on the KK graviton cross section times branching ratio to two photons as a function of the resonance mass with $\coupling=0.1$.
  The cross sections predicted by the benchmark model are computed at LO in QCD using \textsc{Pythia8}, and are shown on the same figure.
  The uncertainty band on the predictions represents the PDF uncertainty estimated from the variations of the NNPDF23LO PDF set.
  The observed limits on the cross section times branching ratio range from \maximumLimitSpinTwoSmallCouplings\ to about \minimumLimitSpinTwoSmallCouplings\ for a KK graviton mass between 500~\GeV\ and 5000~\GeV.
  The RS1 model with $\coupling=0.1$ is excluded for $m_{G^{*}}$ below \limitRS, based on the observed limit determined with pseudo-experiments.
The impact of systematic uncertainties on the expected limit is below 5\% over the entire mass range, and typically 2–-3\% below 2~\TeV.

  \item[] The spin-2 spectrum shown in Figure~\ref{fig:mgg_distributions_spin2} is also interpreted in the context of the ADD model.
A counting experiment is performed in the signal region $\mgg>2240~\GeV$.
In this region, four events are observed in data for $4.3\pm1.0$ expected.
The expected 95\% CL upper limit on the number of excess events is 5.4, and the observed limit has the same value.
The limit on the number of events can be translated into a lower limit on the ultraviolet cutoff scale \ms\ in the ADD model for different theoretical formalisms as summarized in Table~\ref{tab:limit_para}.
Besides the GRW formalism used for the simulation, the Han--Lykken--Zhang (HLZ)~\cite{Han:1998sg} formalism and the Hewett~\cite{Hewett:1998sn} formalism with positive interference are also considered.
A K-factor of about 1.4 was computed using ADD samples generated at LO and NLO using \textsc{MadGraph5\_aMC@NLO},
and is included in the results to indicate the potential impact from the higher-order calculation.
The uncertainty in the signal theory prediction typically varies the limit results by 8\%.
\end{itemize}

\begin{table}[htbp]
 \begin{center}
 {\small
 \begin{tabular}{|c|l||c|c|cccc|} \hline
   & ADD formalism & GRW & Hewett & \multicolumn{4}{c|}{HLZ} \\
   &  Parameter  &    & positive  &  $n = 3$ &$n = 4$ &$n = 5$ &$n = 6$     \\ \hline \hline
   Without K-factor  & $\ms$ observed limit [\TeV]    & 6.8 & 6.1 & 8.1 & 6.8 & 6.1 & 5.7   \\\hline
   With K-factor & $\ms$ observed limit [\TeV]    & 7.2 & 6.5 & 8.6 & 7.2 & 6.5 & 6.1        \\\hline
  \end{tabular}
  }
 \caption{Observed 95\% CL lower limits on the ADD model parameter \ms\ for the GRW, the Hewett (with positive interference) and the HLZ formalisms.
   For the HLZ formalism, the number of extra dimensions ($n$) is varied between 3 and 6.
   The second row shows results based on MC samples generated at LO by \textsc{Sherpa 2.2.1}.
   A K-factor was computed using ADD samples generated at LO and NLO using \textsc{MadGraph5\_aMC@NLO},
   and is included in the results shown in the third row to indicate the potential impact from the higher-order calculation.
 }
 \label{tab:limit_para}
 \end{center}
\end{table}

%% file: sections/conclusion.tex
\section{Conclusion}
\label{sec:conclusion}

Searches for new phenomena in high-mass diphoton final states
with the ATLAS experiment at the LHC are presented. 
The proton--proton collision data corresponding to an integrated luminosity of \intLumi\ were 
recorded in 2015 and 2016 at a centre-of-mass energy of $\sqrt{s}$ = 13~\TeV. Analyses optimized for the 
search for spin-0 resonances with masses above 200~\GeV, for spin-2 resonances predicted by the Randall--Sundrum model with masses above 500~\GeV,
and for non-resonant Kaluza--Klein graviton signals in the Arkani-Hamed--Dimopoulos--Dvali scenario are performed.

The data are consistent with the Standard Model background expectation. At a mass around 750~\GeV, where the largest deviation from the background hypothesis was previously observed, no excess is seen in the 2016 data. In the combined dataset, the largest local deviation from the background-only hypothesis for the spin-0 (spin-2) resonance search is \localSigSpinZero\,$\sigma$ (\localSigSpinTwo\,$\sigma$) for a mass near \bestMassSpinZero~\GeV\ and narrow width (mass near \bestMassSpinTwo~\GeV\ and $\coupling=\bestCouplingSpinTwo$).
The global significance of this excess is null (\globalSigSpinTwo\,$\sigma$) for the spin-0 (spin-2) resonance search.

In the spin-0 resonance search, the observed 95\% CL upper limits on the fiducial cross section times branching ratio for a narrow-width signal range from \maximumLimitSpinZeroNWA\ at 
200~\GeV\ to about \minimumLimitSpinZeroNWA\ at 2700~\GeV.
In the spin-2 resonance search, the observed limits on the cross section times branching ratio for $\coupling=0.1$ range from \maximumLimitSpinTwoSmallCouplings\ to about \minimumLimitSpinTwoSmallCouplings\ for a KK graviton mass between 500~\GeV\ and 5000~\GeV.
The RS1 model with $\coupling=0.1$ is excluded below $m_{G^{*}}=\limitRS$.
These results supersede those previously reported by ATLAS based on 2015 data.

In the ADD scenario, lower limits between \minimumLimitADD\ and \maximumLimitADD\ are set on the ultraviolet cutoff scale \ms,
depending on the number of extra dimensions and the theoretical formalism used.

%% file: acknowledgements/Acknowledgements.tex

We thank CERN for the very successful operation of the LHC, as well as the
support staff from our institutions without whom ATLAS could not be
operated efficiently.

We acknowledge the support of ANPCyT, Argentina; YerPhI, Armenia; ARC, Australia; BMWFW and FWF, Austria; ANAS, Azerbaijan; SSTC, Belarus; CNPq and FAPESP, Brazil; NSERC, NRC and CFI, Canada; CERN; CONICYT, Chile; CAS, MOST and NSFC, China; COLCIENCIAS, Colombia; MSMT CR, MPO CR and VSC CR, Czech Republic; DNRF and DNSRC, Denmark; IN2P3-CNRS, CEA-DSM/IRFU, France; SRNSF, Georgia; BMBF, HGF, and MPG, Germany; GSRT, Greece; RGC, Hong Kong SAR, China; ISF, I-CORE and Benoziyo Center, Israel; INFN, Italy; MEXT and JSPS, Japan; CNRST, Morocco; NWO, Netherlands; RCN, Norway; MNiSW and NCN, Poland; FCT, Portugal; MNE/IFA, Romania; MES of Russia and NRC KI, Russian Federation; JINR; MESTD, Serbia; MSSR, Slovakia; ARRS and MIZ\v{S}, Slovenia; DST/NRF, South Africa; MINECO, Spain; SRC and Wallenberg Foundation, Sweden; SERI, SNSF and Cantons of Bern and Geneva, Switzerland; MOST, Taiwan; TAEK, Turkey; STFC, United Kingdom; DOE and NSF, United States of America. In addition, individual groups and members have received support from BCKDF, the Canada Council, CANARIE, CRC, Compute Canada, FQRNT, and the Ontario Innovation Trust, Canada; EPLANET, ERC, ERDF, FP7, Horizon 2020 and Marie Sk{\l}odowska-Curie Actions, European Union; Investissements d'Avenir Labex and Idex, ANR, R{\'e}gion Auvergne and Fondation Partager le Savoir, France; DFG and AvH Foundation, Germany; Herakleitos, Thales and Aristeia programmes co-financed by EU-ESF and the Greek NSRF; BSF, GIF and Minerva, Israel; BRF, Norway; CERCA Programme Generalitat de Catalunya, Generalitat Valenciana, Spain; the Royal Society and Leverhulme Trust, United Kingdom.

The crucial computing support from all WLCG partners is acknowledged gratefully, in particular from CERN, the ATLAS Tier-1 facilities at TRIUMF (Canada), NDGF (Denmark, Norway, Sweden), CC-IN2P3 (France), KIT/GridKA (Germany), INFN-CNAF (Italy), NL-T1 (Netherlands), PIC (Spain), ASGC (Taiwan), RAL (UK) and BNL (USA), the Tier-2 facilities worldwide and large non-WLCG resource providers. Major contributors of computing resources are listed in Ref.~\cite{ATL-GEN-PUB-2016-002}.

%% file: atlas_authlist.tex
\begin{flushleft}
{\Large The ATLAS Collaboration}

\bigskip

M.~Aaboud$^\textrm{\scriptsize 137d}$,
G.~Aad$^\textrm{\scriptsize 88}$,
B.~Abbott$^\textrm{\scriptsize 115}$,
O.~Abdinov$^\textrm{\scriptsize 12}$$^{,*}$,
B.~Abeloos$^\textrm{\scriptsize 119}$,
S.H.~Abidi$^\textrm{\scriptsize 161}$,
O.S.~AbouZeid$^\textrm{\scriptsize 139}$,
N.L.~Abraham$^\textrm{\scriptsize 151}$,
H.~Abramowicz$^\textrm{\scriptsize 155}$,
H.~Abreu$^\textrm{\scriptsize 154}$
% ...
\bigskip
\\
% ...
$^{12}$ Institute of Physics, Azerbaijan Academy of Sciences, Baku, Azerbaijan\\
% ...
$^{88}$ CPPM, Aix-Marseille Universit{\'e} and CNRS/IN2P3, Marseille, France\\
% ...
$^{115}$ Homer L. Dodge Department of Physics and Astronomy, University of Oklahoma, Norman OK, United States of America\\
% ...
$^{119}$ LAL, Univ. Paris-Sud, CNRS/IN2P3, Universit{\'e} Paris-Saclay, Orsay, France\\
% ...
$^{137}$ $^{(a)}$ Facult{\'e} des Sciences Ain Chock, R{\'e}seau Universitaire de Physique des Hautes Energies - Universit{\'e} Hassan II, Casablanca; $^{(b)}$ Centre National de l'Energie des Sciences Techniques Nucleaires, Rabat; $^{(c)}$ Facult{\'e} des Sciences Semlalia, Universit{\'e} Cadi Ayyad, LPHEA-Marrakech; $^{(d)}$ Facult{\'e} des Sciences, Universit{\'e} Mohamed Premier and LPTPM, Oujda; $^{(e)}$ Facult{\'e} des sciences, Universit{\'e} Mohammed V, Rabat, Morocco\\
% ...
$^{139}$ Santa Cruz Institute for Particle Physics, University of California Santa Cruz, Santa Cruz CA, United States of America\\
% ...
$^{151}$ Department of Physics and Astronomy, University of Sussex, Brighton, United Kingdom\\
% ...
$^{154}$ Department of Physics, Technion: Israel Institute of Technology, Haifa, Israel\\
$^{155}$ Raymond and Beverly Sackler School of Physics and Astronomy, Tel Aviv University, Tel Aviv, Israel\\
% ...
$^{161}$ Department of Physics, University of Toronto, Toronto ON, Canada\\
% ...
$^{a}$ Also at Department of Physics, King's College London, London, United Kingdom\\
$^{b}$ Also at Institute of Physics, Azerbaijan Academy of Sciences, Baku, Azerbaijan\\
$^{c}$ Also at Novosibirsk State University, Novosibirsk, Russia\\
$^{d}$ Also at TRIUMF, Vancouver BC, Canada\\
$^{e}$ Also at Department of Physics {\&} Astronomy, University of Louisville, Louisville, KY, United States of America\\
% ...
$^{*}$ Deceased
\end{flushleft}
